\documentstyle[multicol,epsfig,aps,prb,array]{revtex}

\def\dfrac#1#2{{\displaystyle{#1\over#2}}}

\begin{document}

\title{General boundary conditions for the  envelope function in
multiband ${\bf k} \cdot {\bf p}$ model.}

\author{A.~V.~Rodina\cite{Ioffe}}

\address{I.~Physics Institute, Justus Liebig University of Giessen,
35392 Giessen, Germany}

\author{A.~Yu.~Alekseev}

\address{Institute of Theoretical Physics, Uppsala University,
S-75108, Uppsala, Sweden }

\address{Department of Mathematics, University of Geneva, 1211
Geneva, Switzerland}

\author{Al.~L.~Efros and M.~Rosen}

\address{Naval Research
Laboratory, Washington DC 20375, USA}

\author{B.~K.~Meyer}

\address{I.~Physics Institute, Justus Liebig University of Giessen,
35392 Giessen, Germany}


\maketitle

\begin{abstract}

We have derived general   boundary  conditions
(BC) for the multiband envelope functions (which do not contain
spurious solutions) in  semiconductor heterostructures  with abrupt
heterointerfaces. These  BC require the conservation of the
probability flux density normal  to the interface and guarantee that
the multiband Hamiltonian  be self--adjoint.  The BC are energy
independent and are characteristic properties of the interface.
Calculations have been performed of the effect of the general BC  on
the electron energy levels in a potential well with infinite
potential barriers  using a coupled two band model.  The connection
with other approaches to determining BC for the envelope function
and to the spurious solution problem  in the multiband ${\bf k}
\cdot {\bf p}$ model are discussed.

\pacs{PACS numbers:
73.21.-b; 73.22.-f; 73.20.-r; 02.30.Nw}

\end{abstract}

\section{Introduction}

It is impossible to overestimate the role of
the multiband effective mass approximation  (MEMA), which consists
of the multiband   ${\bf k} \cdot {\bf p}$ method together with the
envelope function approximation (EFA), in the simulation of
electronic and optical devices formed from various  semiconductor
heterostructures, such as those with Type-I and Type II
heterointerfaces,  quantum well structures and  superlattices, and
one and zero dimensional semiconductor structures such as quantum
wires and quantum dots. The calculations of energy bands and wave
functions in such structures  have often been considered as only
being a matter of "energy band engineering" or "wave function
engineering".  However, although such calculations successfully
describe many heterostructures (see, for example, Ref.
\onlinecite{Ram-Mohan}),  a general theoretical description of
structures with abrupt interfaces  has  yet to be developed.

The MEMA gives an accurate description of the energy band structure
of bulk semiconductors near the extrema of the first Brillouin zone.
In this approach the carrier wave functions are expanded
 in the band edge Bloch functions near an extremum with
"envelope function" coefficients, or components, that are required
to vary slowly over a distance  on the scale of the unit cell
$a_0$.  The bulk MEMA  Hamiltonian contains such parameters    as
band edge energies, interband momentum matrix elements and carrier
effective masses.  These material parameters  are determined by
fitting to independent experimental data  or are obtained from first
principles calculations. The  MEMA  is certainly appropriate for
calculating the spectrum and carrier wave functions in the  presence
of  any slowly varying potential \cite{birpikus}.

However, abrupt heterointerfaces in semiconductor heterostructures
break the periodicity of the crystal potential
\cite{ivchenko,tamm}.  The  nonperiodic part of the potential varies
rapidly in an interface region of size $a \approx a_0$. This may
lead to such interface effects as the admixture of remote  band
states \cite{ivchenko}, coupling of states with different symmetry
\cite{andoII,fu,ivkamros,foremanburt}, and the   formation of
surface localized states \cite{tamm}.  Moreover, Bloch functions  in
dissimilar materials can be significantly different \cite{einevoll}
and do not even exist in the boundary regions \cite{laikhtman}. In
light of the great success of the MEMA, it is important  to  develop
an adequate description of semiconductor heterostructures with
abrupt boundaries using the MEMA to describe each material in the
heterostructure. Laikhtman \cite{laikhtman} has shown that  this is
possible for heterostructures with a  characteristic length $L \gg
a$, provided that the boundaries are treated appropriately. The
problem is then focused on the appropriate choice of boundary
conditions (BC) for piece--wise defined envelope functions in these
heterostructures.

Conservation of the  current normal to the interface is an obvious
requirement dictated by conservation of probability density.  In the
EFA this condition reduces to   the continuity across the interface
of the normal component of the probability flux density $J_\tau$
averaged over the unit cell \cite{ivchenko,gelmont}.  In a
heterostructure formed of semiconductors with parabolic bands, which
can be described by  a single band effective mass approximation
(EMA) (the simplest MEMA), this condition can be expressed as the
continuity across the interface of:
\begin{equation} J_\tau= - i \frac{\hbar}{2m} \left(
f^{*} f^{'} - {f^{'}}^{*} f \right) , \label{fluxa}
\end{equation}
  where  $\vec{\tau}$ is a unit vector directed
normal to the interface, $m$ is the effective mass, $f$ is the
single band envelope function and the prime denotes the
derivative along $\vec{\tau}$. The continuity of Eq. (\ref{fluxa})
is satisfied  if
\begin{equation} f_{\it r} = f_l, \qquad
\frac{f^{'}_{\it r}}{m_{\it r}}  = \frac{f^{'}_l}{m_l} \, ,
\label{flr} \end{equation}
where the indices $r$ and $l$ refer to a
point just to the right and just to the left, respectively,  of the
interface. These BC, expressing the separate continuity of the envelope
function and the normal envelope velocity, $f^{'}/m$, first
suggested  in Refs. \onlinecite{conley,duke}, are  referred to as
"conventional" BC.  However, the conventional BC of Eq. (\ref{flr})
were recognized almost 20 years ago \cite{mori} to be a special case
of more general BC that conserve $J_\tau$ across the interface
\cite{ivchenko,laikhtman,mori,andoI,balian,pistol}:
\begin{equation}
\left ( \begin{array}{c}
f_{\it r} \\ f^{'}_{\it r}/m_{\it r}
\end{array} \right) =  T_{tr} \left(
\begin{array}{c}
f_l \\ f^{'}_l/m_l
\end{array} \right)\, , \qquad   \,
\det \left[ T_{tr} \right] = t_{11}t_{22}- t_{12}t_{21} = 1 \label{tlr}
\end{equation}
 where the elements, $t_{ij}$, of the transfer matrix, $T_{tr}$,
 are  characteristic of the
interface\cite{laikhtman,ivchenko}, and, as seen from a more general
derivation of Eq. (\ref{tlr}) \cite{balian,pistol}, based on the
requirement for  the EMA Hamiltonian to be self--adjoint
\cite{reed}, do not depend on the particular states and are energy
independent. For example, first principle calculations performed for
$GaAs/Al_xGa_{1-x}As$ heterostructures \cite{andoI} show that the
off diagonal  elements $t_{12}$ (which was shown to be always
proportional to the small parameter $a/L$ \cite{laikhtman}) and
$t_{21}$ are negligible for this interface. In this case the BC can be
written in the form
\begin{equation}
m^{\alpha}_r f_r = m^{\alpha}_l f_l, \qquad \frac{f^{'}_r}{m^{\alpha
+1}_r} = \frac{f^{'}_l}{m^{\alpha +1}_l} \, ,
\label{flra}
\end{equation}
similar to the general BC first suggested in Ref.
\onlinecite{harrison}. In general the parameters $\alpha$ and the
off diagonal element $t_{21}$ (which may be non-zero for other
heterostructures \cite{kroemer}) strongly  effect
the penetration of the wave  function into the barrier  and  the
energy spectra in quantum wells (QWs). For example, this  has
been demonstrated in Ref. \onlinecite{SST95} by analyzing the
experimental data for the exciton optical spectra in narrow
$GaAs/AlGaAs$ and $InGaAs/InP$ QWs using the EMA with modified BC
similar to those given by Eq. (\ref{flra}). However, the parabolic
band approximation has a very narrow range of applicability and
appropriate BC for the MEMA need to be developed for describing
experimental data.

For the multiband analysis of heterostructures, a natural extension
of the conventional BC of Eq. (\ref{flr}) is the  continuity of each
component of the $N$--component envelope function, $\Psi$, and of
each component of the normal envelope velocity, $V_\tau$, at each
interface.\cite{bastard8182,alt83,baraff} Taken together, these BC
preserve the normal component of the envelope flux density,
$J_\tau$, averaged over a unit cell. The most general derivation of
these BC, suggested by Barraff and Gershoni,\cite{baraff} is based
on the requirement for the MEMA Hamiltonian to be self-adjoint.  A
general form of the BC within the  MEMA for the planar
heterostructures was suggested by Kisin {\em et al.} \cite{gelmont}.
Those authors start from the conservation of $J_\tau$ at the
interface, and introduce a transfer matrix connecting the components
of the multiband envelope function, $\Psi$, and its normal
derivative, $\Psi^{'}$,  on each side of the heterointerface. The
general consideration of the $8 \times 8$ band model   based on the
symmetrical properties of the planar heterointerface \cite{gelmont}
has allowed us to gain important information about the elements of
the transfer matrices. However, the BC introduced in Ref.
\onlinecite{gelmont}\,  may lead to an energy dependent connection
between  the components of $\Psi$ and  $V_\tau$ on the left and
right sides of the heterointerface, and, therefore, do not generally
satisfy the requirement that  the MEMA Hamiltonian be
self-adjoint.

The MEMA  that take band coupling into account have a
significant theoretical problem when describing various semiconductor
heterostructures. The full expansion of the $N$-band envelope
function  in the $N$ independent plane wave solutions of the bulk
Hamiltonian corresponding to the same energy $\varepsilon$ includes
solutions with a very large wave vector lying outside the first
Brillouin zone \cite{white,schuur}. Such  "wing" (evanescent)
\cite{white} or "spurious" (oscillatory) \cite{schuur} solutions
must be rejected in the perturbation ${\bf k} \cdot {\bf p}$
theory; in the bulk they are rejected as being
unphysical solutions  \cite{witold,godfrey,zung}. However,  their
elimination from the multiband envelope function for
heterostructures \cite{winkler,meney,godfrey,gelmont} leads to
additional complications in the analysis of the BC. For example, the
continuity of all $N$ components of this truncated envelope
function  can not be satisfied if the interband  matrix
elements of the momentum operator $\hat {\bf p}$ are not continuous
 across the interface \cite{gelmont}. The
same problem remains for the alternative approach,  based on
modifying the MEMA Hamiltonian by discarding those terms responsible
for the large $k$ solutions \cite{white,schuur,eppenga,foreman}. The
use of different BC for the conduction and valence band states
(see, for example,  Ref. \onlinecite{eppenga}) automatically results
in the dependence of the BC on $\varepsilon$, and is justified only
if band coupling in the MEMA Hamiltonian and bands
nonparabolicity can be neglected.

In our paper we consider the boundary condition problem within the
MEMA for heterostructures with abrupt boundaries.  We show that  the
most general requirement for the MEMA Hamiltonian to be  self--adjoint
 implies continuity at the interface of the normal component of the
envelope flux density vector $J^{\alpha\beta}_\tau$.
 Here the indices $\alpha$ and $\beta$ denote, in general, two
arbitrarily chosen envelope functions $\Psi_\alpha$ and
$\Psi_\beta$, respectively. This  allows us to formulate general BC
using a new {\em energy and state independent\,} transfer matrix
connecting the components of the envelope functions and their normal
velocities on each side of the interface. It also allows us to find
a general  way of eliminating the large--$k$ solutions because these
solutions make only a small (proportional to $a/L$) contribution to
the envelope flux density. This procedure does not depend on the
particular nature of the large-$k$ solutions (which we shall
hereafter call spurious) and can be applied to boundaries with
either finite or infinite potential barriers. The elements of these
transfer matrices can be considered  as characteristic surface or
interface parameters and are additional to the bulk energy band
parameters in the multiband  ${\bf k} \cdot {\bf p}$ approximation.

The paper is organized as follows.  In Sec. \ref{efabc}\,  we derive
the general form of BC for the  multiband envelope function, that
follows from the condition that the  MEMA Hamiltonian be
self--adjoint. In Sec. \ref{spur} we derive a truncated form of
general BC for slowly varying envelope functions which do not allow
spurious solutions. To clarify the general procedure, we first
consider, in detail,
 the two band Kane model. Sec. \ref{calc} gives analytical and
numerical examples of the effect of the general BC on the quantum
size energy levels in dots with infinite potential barriers. In Sec.
\ref{compar} we discuss our results and compare them with those of
other approaches to the BC problem within the MEMA and other ${\bf
k} \cdot {\bf p}$ theories for heterostructures .

\section{General form of multiband Boundary Conditions in
heterostructures with abrupt boundaries}
\label{efabc}

Let us consider a semiconductor heterostructure  made of $M$
arbitrary shaped semiconductor layers with a characteristic length
$L$. Single--electron wave functions,  ${ \Phi}({\bf r})$, in these
structures are  solutions of the Schr\"{o}dinger equation with the
microscopic  Hamiltonian $\hat H_{\rm micr}$ containing a crystal
potential. This potential is  "periodic" inside each of the
bulk--like semiconductor layers,  but it varies  rapidly in the
boundary regions with width  $a \approx a_0$. Assuming that $a/L$ is
small, we neglect the exact behavior of $\Phi({\bf r})$ in  the
boundary regions and  expand $\Phi^j \equiv \Phi({\bf r} \in {\bf
R}_j)$ in the bulk--like interior region ${\bf R}_j$ of the $j$-th
layer  in the Bloch functions, $u^j_{nk_{0j}}$, at a critical point,
$k_{0j}$, of the bulk
 energy band structure for the material in that layer:
\begin{equation}
{ \Phi^j ({\bf r})} =
 \sum_{n=1}^{N_j} \Psi_{n}^j({\bf r}) u^j_{nk_{0j}} \, , \qquad j=1,2\dots
 M
\label{psi1}
\end{equation}
where $N_j$ is the number of bands that  describe
  the band-edge bulk properties of the material in the $j$-th
layer,  and $M$
 is the number of layers.  The symmetry of the material and the
number of bands can be different in each region. The $N_j$
component envelope function
 ${\Psi}^j({\bf r})=\{ \Psi_n^j({\bf r})\}_{n=1}^{N_j}$ is
 slowly varying in the bulk--like region ${\bf r}\in {\bf R}_j$,
 where it satisfies the  ${\bf k} \cdot {\bf p}$ Schr\"{o}dinger
 equation
  \begin{equation}
\hat H^j (\hat {\bf k})  { \Psi}^j({\bf r})= \varepsilon { \Psi}^j
({\bf r}) \,~,~{\bf r}\in {\bf R}_j \, .
 \label{Hamjj}~
\end{equation}
 The $N_j \times N_j$ Hamiltonians $\hat H^j$ are obtained after
averaging of the perturbed microscopic Hamiltonian $\hat H_{\rm
micr}(\hat{\bf p} + \hbar \hat{\bf k})$ over a unit cell in each
bulk--like region \cite{birpikus}  and includes terms up to the
second order in the wave vector operator $\hat{\bf k} = -i {\bf
\nabla}$. The  boundary regions, however, are excluded from the
averaging procedure, and the parameters of the resulting
heterostructure MEMA Hamiltonians,
$\hat H^j$, may have abrupt jumps from one layer
to another.

In what follows we study in details the boundary conditions imposed
 on the wave functions $\Psi({\bf r}) \equiv \Psi^j({\bf r})$
for ${\bf r} \in {\bf R}_j$  at the interfaces. The microscopic
Hamiltonian $\hat H_{\rm micr}$ acting on $\Phi({\bf r})$ is
 a self--adjoint operator \cite{reed}. This is
equivalent to the conservation of charge in the
full heterostructure region $\Omega$:
\begin{equation}
\frac{d}{dt} \int_{\Omega} {\bf d}^3 {\bf r}\, |\Phi|^2\, = 0\, .
\label{ddt}
\end{equation}
If  $a/L$ is small, the contribution from the boundary regions can
be neglected. Therefore,  Eq. (\ref{ddt}), within
the accuracy of the MEMA, is replaced by $d/dt \left< \Psi,\Psi
\right> \, =0$, where $\Psi$ is the corresponding envelope function
satisfying the BC and the inner product is now defined as
\begin{eqnarray}
\left<  { \Psi}_\alpha \, , { \Psi}_\beta \, \right> = \sum_{j}
\int_{{\bf R}_j}  {\bf d}^3 {\bf r}\, \left(   { \Psi}^j_\alpha, {
\Psi}^j_\beta    \right ) \, , \qquad
 \left(   {
\Psi}^j_\alpha, { \Psi}^j_\beta    \right ) =
 \sum_{n=1}^{N_j}  { \Psi}^{j \,
*}_{\alpha \, n} {\Psi}^j_{\beta \, n} .
 \label{inner}
\end{eqnarray}
This implies, in turn, that the
 heterostructure MEMA Hamiltonian, $\hat H \equiv \hat H^j$ for
 ${\bf r} \in {\bf R}_j$, acting on $\Psi$
has to be a self--adjoint operator. Therefore for any two arbitrarily
chosen functions $\Psi_\alpha$ and $\Psi_\beta$, the condition
\begin{eqnarray}
 \left<  \hat H {\Psi}_\alpha,  { \Psi}_\beta    \right > =
\left<  { \Psi}_\alpha,  \hat H  { \Psi}_\beta    \right >
   \label{self}
\end{eqnarray}
must be satisfied if the same  BC is imposed on $\Psi_{\alpha}$ as
 well as on  $\Psi_\beta$ (see, for example, Ref. \onlinecite{peres}).
  The MEMA Hamiltonians $\hat H^j$, when defined  for the ideally
 infinite homogenous bulk semiconductor, are self--adjoint, but when
one adds the abrupt boundaries, the condition of Eq. (\ref{self})
may be violated by boundary terms. The requirement for $\hat H$ to
be self--adjoint limits the choice of the boundary conditions that
may occur, but there is an ample variety of BC that satisfy it.  We
shall see   that any self--adjoint MEMA Hamiltonian  preserves the
normal component of the  envelope flux density for an arbitrary
state of the system.

 The MEMA Hamiltonians  for the $N_j$
component envelope wave functions have terms of first and  second
order in $\hat {\bf k}$:
\begin{equation}
\hat   H^j = \hat C^j + \hbar \hat {\bf B}^j_\mu\hat{\bf k}_\mu  +
\hbar^2  \hat D_{\mu \, \nu}^j \hat k_\mu  \hat k_\nu \, ,
\label{Nham}
\end{equation}
where  $\hat C^j$, each $\hat{B}_\mu^j$ ($\mu = x,y,z$) and $\hat
D_{\mu \, \nu}^j$ ($\mu \, , \nu = x,y,z$) are Hermitian  $N_j
\times N_j$ tensors of rank 0,1, and 2, respectively. They contain
the energy band parameters for the material in ${\bf R}_j$ and are
defined only in these bulk--like regions. Using Eq. (\ref{Nham}) we
can write a  velocity operator $\hat {\bf V} \equiv \hat {\bf V}^j$
for ${\bf r} \in {\bf R}_j$ which acts on the heterostructure
envelope functions:
\begin{equation}
\hat {\bf V}^j = \frac{1}{\hbar} \frac{\partial \hat H^j}{\partial
{\bf k}} =   \hat {\bf B}^j + \hbar \frac{\partial \hat D^j_{\mu \,
\nu} \hat k_\mu  \hat k_\nu }{\partial {\bf  k}}  \, .
\label{efavelocity}
 \end{equation}
A general envelope  flux density matrix can be
written:
\begin{equation}
{\bf J}^{\alpha \, \beta}({\bf r}) =  \frac{1}{2} \left[ \left(
{\Psi}_\alpha, \hat {\bf V}{\Psi}_\beta \right) + \left( \hat {\bf
V}{\Psi}_\alpha , {\Psi}_\beta\right) \right] \, , \quad {\bf r} \in
{\bf R}_j \, ,
\end{equation}
where ${ \Psi}_\alpha$ and ${ \Psi}_\beta$  are two arbitrarily
chosen   solutions ${\Psi}_\alpha^j \equiv {\Psi}_\alpha({\bf r} \in
{\bf R}_j)$, ${\Psi}_\beta^j \equiv {\Psi}_\beta({\bf r} \in {\bf
R}_j)$ of the Schr\"{o}dinger equation Eq. (\ref{Hamjj}),  defined
in each bulk--like region, with energies $\varepsilon_\alpha$ and
$\varepsilon_\beta$ respectively. Noting that
\begin{eqnarray}
{\rm div} {\bf J}^{\alpha \, \beta}({\bf r}) = - \frac{i}{2} \left[
(\hat {\bf k}  { \Psi}_\alpha, \hat {\bf V} { \Psi}_\beta) - ( {
\Psi}_\alpha, \hat {\bf k} \hat {\bf V} {\Psi}_\beta) - \right.
\nonumber \\ \left. ( \hat {\bf V} { \Psi}_\alpha, \hat {\bf k} {
\Psi}_\beta ) + ( \hat {\bf k} \hat {\bf V} {\Psi}_\alpha, {
\Psi}_\beta ) \right] \, ,  \quad {\bf r} \in {\bf R}_j \,
,\nonumber
\end{eqnarray}
and using Eqs. (\ref{Nham}, \ref{efavelocity})
we arrive at the identity:
\begin{equation}
\left( {\Psi}_\alpha,  \hat H { \Psi}_\beta \right) - \left( \hat H
{ \Psi}_\alpha ,  { \Psi}_\beta \right)  = \frac{\hbar}{i} \, {\rm
div} {\bf J}^{\alpha \, \beta}({\bf r}) \, , \quad {\bf r} \in {\bf
R}_j \, . \label{ident}
 \end{equation}
Thus, for the MEMA Hamiltonian $\hat H$  to be self--adjoint with
BC imposed on $\Psi_\alpha$ and $\Psi_\beta$, Eq. (\ref{self}) requires
\begin{equation}
\sum_j \int_{{\bf R}_j} d^3{\bf r} \cdot \,  {\rm div} \, {\bf
J}^{\alpha \, \beta} = \sum_j  \int_{{\bf R}_j} d {\bf S}_j \cdot
{\bf J}^{ \alpha \, \beta} = 0 \, ,
 \label{sumintflux}
\end{equation}
where ${\bf dS}_j$ is an element of the surface bounding ${\bf
R}_j$. Noting that Eq. (\ref{sumintflux}) holds for arbitrary
surface shape, we find that local continuity of the normal envelope
flux density matrix across each interface must hold:
\begin{equation}
J_{\tau}^{\alpha \, \beta}=\vec{\tau} \cdot {\bf
J}^{\alpha \, \beta} = \frac{1}{2}  \left[ \left( {\Psi}_\alpha,
 \vec{\tau}  \cdot \hat {\bf V}
{\Psi}_\beta \right) + \left( \vec{\tau} \cdot \hat {\bf V}
{\Psi}_\alpha , {\Psi}_\beta \right) \right] = const \, .
\label{generalflux}
\end{equation}
Equation (\ref{generalflux})  is a generalization of the normal flux
density conservation law, $J_\tau\equiv J_{\tau}^{\alpha \,
\alpha}=const$, that is often used as the starting point for
deriving BC\cite{ivchenko,gelmont}.  The condition  $J_\tau^{\alpha
\beta}=const$ is a very general and strong  requirement on the
envelope functions at each point of the  interface. For an
unpenetratable  interface, or a semiconductor/vacuum
 surface it reduces to $J_\tau^{\alpha \beta}=0$. We will consider
this case only in Section \ref{calc}, and shall assume, for now,
that $J_\tau^{\alpha \beta} \ne 0$ at  the semicondutor
heterointerface.

The condition of Eq. (\ref{generalflux}) must be fulfilled
independently of whether the number of the envelope function
components on the two sides of interface is the same. For clarity in
the following we will  only consider heterostructures formed of
semiconductors whose energy band structures are described by
multiband Hamiltonians $\hat{H}^j$ of same symmetry and with $N_j=N$
in each region. For the more general case, a more complicated
procedure similar to that suggested in Ref. \onlinecite{laikhtman}
is needed.   The appropriate number of boundary conditions  depends
on the number of independent components of the envelope function
${\Psi}$ and normal envelope velocity  $ V_{\tau}={\bf \tau} \cdot
{\bf V} {\Psi}$.
  If all the components of
${\Psi}$ and ${V_\tau}$,  are linearly independent on each side of
the interface, the most general BC
one can impose on  the envelope functions are of the form:
\begin{equation}
\left ( \begin{array}{c}
{ \Psi}^{+} \\ i {V_{\tau \, }^{+}}
\end{array} \right) = {T}_{2N} \left(
\begin{array}{c}
{ \Psi}^{-} \\ i { V_{\bf \tau \, }^{-}}
\end{array} \right) \, , \label{matrixab}
\end{equation}
where  the  indices "-" and "+"  refer to neighboring points on
neighboring bounding surfaces  $S_j$ and $S_{j+1}$, respectively,
and $T_{2N}$ is the $2N \times 2N$ transfer matrix.
In the BC described by Eq.(\ref{matrixab}) the material
 parameters of the MEMA Hamiltonian are included in the vectors
 of the normal envelope velocity.
 Equation
(\ref{generalflux}) is satisfied for all arbitrarily chosen $\alpha$
and $\beta$ if and only if the BC of Eq. (\ref{matrixab})
 are imposed on  all $\Psi_\alpha$  as well as on all  $\Psi_\beta$
 with the same $T_{2N}$ and
 \begin{equation}
\qquad {T}^{\dagger}_{2N} \left(  \begin{array}{c c}
0 & I_N \\ - I_N & 0
\end{array} \right) {T}_{2N} = \left(  \begin{array}{c c}
0 & I_N \\ - I_N & 0
\end{array} \right),
\label{matrixgen}
\end{equation}
where  $I_N$ is the $N \times N$ unit matrix and the dagger denotes
the Hermitian conjugate. Therefore,
 $T_{2N}$ must be state independent and characteristic only of the
interface. For   the conventional BC, $ { \Psi}^+ =  { \Psi}^- \, ,
{ V}_{{\tau}}^+ ={V}_{{\tau}}^- \, ,$ the  transfer matrix reduces
to the unit matrix,  $T_{2N}=I_{2N}$.

In the "pure" Kane models ($\hat D_{\mu \, \nu} \equiv 0$  and
$\hat{H}$ only contains terms linear in $\hat {\bf k}$) the $N$
components of $V_{\tau}=\vec{\tau}  \cdot \hat {\bf B} \Psi$ depend
linearly  on the $N$ components of $\Psi$.  In this case, the
general BC can be written
\begin{equation}
{\Psi}^+ = T_{N} {\Psi}^- \,  , \qquad  T_{N}^{\dagger} B_{\tau}^ +
T_{N} =  B_{\tau}^- \, . \label{transf1}
\end{equation}
Here $B_{\tau}=
\vec{\tau} \cdot \hat {\bf B}$ and $T_{N}$ is an $N \times N$ state
 independent transfer matrix.
One can see that if $B_{\tau}^+ \ne B_{\tau}^-$, the transfer matrix
$T_{N}$ can not be the unit matrix $I_N$.  The number of independent
equations in Eq. (\ref{transf1}) is determined by  the rank of the
matrices $B_\tau$ and can be less then $N$.

The elements of the state independent transfer matrix  may depend
on such properties of the heterojunction as band offsets and surface
 crystal symmetry due to, e.g., its
reconstructions. Additional restrictions on the components of the
transfer matrix can be obtained from symmetry consideration of the
bulk and  surface properties \cite{gelmont,ivkamros} in  particular
cases. For heterostructures which geometrical shape has some
elements of symmetry one can expect the  boundary parameters to be
the same at the symmetrical points of the interface. For example, in
a QW the parameters of the transfer matrix are the same along the
plane, and in the spherically shaped heterostructure, the parameters
of the transfer matrix are the same anywhere on the surface.

\section{Elimination of spurious solutions from multiband ${\bf k}
\cdot {\bf p}$ model: truncated  Boundary Conditions}
\label{spur}

Consider a heterostructure with a single  heterointerface described
by a piece-wise  $N \times N$ Hamiltonian of Eq. (\ref{Nham}), and
with nonzero  energy dispersion matrices $D_{\mu \nu}$.  The
envelope functions can be expanded in a complete set of $N$ plane
wave eigenfunctions,  ${\Psi}_i(k_i)$, of the $N \times N$ MEMA
Hamiltonian:
\begin{equation}
{\Psi}=\sum_{i=1}^NC_i{\Psi}_i(k_i)~,
\label{expention}
\end{equation}
where the $N$ values of $k_i^2$ are the roots of the  bulk energy
dispersion equation $\det[H(k)-\varepsilon]=0$, and $C_i$ are
numerical coefficients. The set of $2N$ boundary condition equation
in Eq. (\ref{matrixgen}) can be solved only if the expansion  in Eq.
(\ref{expention}), for both sides of the interface, includes all $N$
of the plane wave eigenfunctions. If any, say $m$ roots $|k_i|$
($i=N - m+1 \, ,\dots N$)  lie outside the first Brillouin zone, the
corresponding spurious solutions  must be excluded from  the wave
function. In this case, $2 m$ components of the envelope wave
function and  their velocities are linearly dependent on the other
components, and only $2(N-m)$ independent BC can be imposed on the
truncated function $\tilde { \Psi}=\sum_{i=1}^{N-m}C_i
{\Psi}_i(k_i)$.  In this section we suggest an approximate procedure
which allows us to derive a truncated set of general BC that must be
imposed on the envelope function in order to satisfy
$J_\tau^{\alpha\, \beta}=const$. For the sake of clarity, we first
consider  a two band  model ($N=2$, $m=1$) and a planar
heterointerface. We  then generalize it to arbitrary $N$ and $m$ and
an arbitrarily shaped heterointerface.

The  Hamiltonian of the two band Kane model is written:
\begin{equation}
\hat H=  \left(
\begin{array}{cc}
E_{c} &  0\\ 0 & E_{v}
\end{array} \right) + \hat H_K \, ,
\qquad \hat H_K=  \frac{\hbar^2}{2 m_{0}}\left(
\begin{array}{cc}
\alpha_c \hat k_z^2&  i
 \dfrac{2 P}{\hbar}  \hat k_z \\ -i \dfrac{2 P}{\hbar} \hat k_z &
- \alpha_v \hat k_z^2
\end{array} \right) \, ,
\label{KanedH}
\end{equation}
where $\hat k_z = -i {d}/{dz}$ is the wave vector along the normal
to the interface, $m_0$ is the free electron mass, $P$ is the Kane
interband matrix element, $\alpha_c$ and $\alpha_v$ describe the
contribution of the remote bands to the electron, $m_c$, and hole,
$m_v$, effective masses respectively, and the distance between the
bottom of conduction band, $E_c$, and the top of the valence band,  $E_v$,
is the energy gap $E_{g}=E_{c} - E_{v}$. The  band edge
effective masses can be written as  $m_0/m_{c,v}= \alpha_{c,v} +
E_p/E_g$.  Here  $E_p=2P^{2}/m_0$ is the Kane energy which also
characterizes the nonparabolicity of the electron and hole  bulk
energy spectra. Although this two band model completely  neglects
spin-orbit interactions,  it  describes the  electron and light hole
band coupling in real semiconductors and is often used  in one
dimensional  heterojunctions \cite{gelmont} and spherical dots
\cite{vahala,zung,sercel}. For most semiconductors, $|\alpha_c|\sim
|\alpha_v|\sim 1$ (they can be either positive or negative) and
$E_p\approx 20\pm 5$\,eV. In the "pure" Kane model
$\alpha_c=\alpha_v=0$.

For a planar heterointerface, the general normal  flux density continuity
condition, Eq. (\ref{generalflux}), takes the form:
\begin{equation}
J_\tau^{\Psi\, \Phi}(z)=\frac{1}{2} \left[ \left( {\Psi},   V_{\Phi}
\right) + \left(V_{\Psi}, { \Phi} \right) \right]= const \, ,
\label{selffluxdk}
\end{equation}
where $\Psi$ and $\Phi$ are two arbitrary chosen eigenfunctions of
the Hamiltonian in Eq. (\ref{KanedH}),   ${V}_{\Psi}= \hat V_z
{\Psi}$ and ${V}_{\Phi}= \hat V_z {\Phi}$  are the respective normal
velocities calculated with  the two band Kane model velocity
operator $\hat V_z$
\begin{equation}
{\hat V_z}=   \frac{\hbar}{m_0}  \left(
\begin{array}{cc}
  \alpha_c \hat k_z & i \dfrac{P}{\hbar}    \\
- i \dfrac{P}{\hbar} & - \alpha_v \hat k_z
\end{array} \right).
\label{velocitydH}
\end{equation}
 The general BC, Eqs. (\ref{matrixab}, \ref{matrixgen}), now reduce to
\begin{equation}
\left ( \begin{array}{c}
{ \Psi}^{\it r} \\ i { V_\Psi^{\it r}}
\end{array} \right) = {T}_{tr} \left(
\begin{array}{c}
{ \Psi}^{l} \\ i {V_\Psi^{l}}
\end{array} \right)
, \qquad
{T}^{\dagger}_{tr} \left(  \begin{array}{c c}
0 & I_2 \\ - I_2 & 0
\end{array} \right) {T}_{tr} = \left(  \begin{array}{c c}
0 & \hat I_2 \\ - \hat I_2 & 0
\end{array} \right)\, ,
\label{matrixd}
\end{equation}
where  $T_{tr}$ is an energy independent $4\times 4$ transfer
matrix, and $r$ ($l$) refers to a point to the right (left) side of
 the planar interface. The conventional BC ($\Psi^{\it r}=\Psi^l$ and
$V_\Psi^{\it r}=V_\Psi^l$) are described by the unite transfer
matrix $T_{tr} \equiv I_4$.  In the general case, $\alpha_c \alpha_v
\ne 0$, the BC of Eq.   (\ref{matrixd})  represent 4 linear
independent equations for two (conduction and valence band) wave
function components $\Psi_c$, $\Psi_v$ and two normal velocity
components $V_{\Psi \, c}$, $V_{\Psi \, v}$. In order to solve them,
the envelope function of a state with energy $\varepsilon$ is
written as a linear superposition of two plane wave eigenfunctions
of the Hamiltonian, Eq.(\ref{KanedH}):  ${\Psi}  = C_1{\Psi}_1(k_1)
+ C_2{ \Psi}_2(k_2)$, where $k_{1,2}^{2}$ are the solutions of the
bulk energy  dispersion equation
 \begin{equation}
\left( \frac{\hbar^2 k^2}{2 m_0} \alpha_c + E^{j}_c -  \varepsilon
\right) \left( \frac{\hbar^2 k^2}{2 m_0} \alpha_v - E^{j}_v +
\varepsilon \right) + \frac{\hbar^2 k^2}{2 m_0} E^{j}_p =0.
\label{dispersion}
\end{equation}
Generally we are only interested in states whose energy is either on
the order of or smaller than the semiconductor energy gap.  Using
the condition  $E_p/|\alpha_c \alpha_v| \gg  E_g, |\varepsilon -
E_c|, |\varepsilon - E_v|$, we find for the two roots of the
dispersion equation:
\begin{equation}
\hbar^2 k_1^{2} \approx \frac{2 \left( \varepsilon - E_v\right)
\left( \varepsilon -E_c \right) m_0 }{ \left(\varepsilon -
E_v\right)\alpha_c + \left(E_c -\varepsilon \right)\alpha_v +
E_p } \, , \qquad \hbar^2 k_2^{ 2} \approx -\frac{2E_p
m_0}{\alpha_c \alpha_v} \, .
\label{kroots}
\end{equation}
For typical values of the bulk energy band parameters,  the second
wave vector $|k_2|\sim 1\,  \AA^{-1}$  lies beyond the first
Brillouin zone in which the MEMA is valid.  The corresponding
eigenvector, ${\Psi}_2(k_2)$ is a spurious  solution (either
evanescent or propagating, depending on the sign of the product
$(\alpha_c \alpha_v)$ \cite{witold})  and  must be excluded from the
expansion of the total wave function.  The continuity  of the normal
flux density, Eq. (\ref{selffluxdk}), is now limited to slowly
varying functions of $k_1$, $\Psi= {\Psi}_1(k_1)$  and  $\Phi =
{\Phi}_1(k_1)$, and leads to $J_{\tau, 1}^{\Psi \, \Phi}(z)\equiv
J_{\tau}^{\Psi_1 \, \Phi_1}(z)$:
\begin{eqnarray}
& & J_{\tau, 1}^{\Psi \, \Phi}(z) = \frac{i \, P}{m_0} \left[ \Psi^{*}_{c}
 \Phi_{v} \left( 1 + \alpha_c \frac{\varepsilon_\Phi - E_v}{E_p} +
 \alpha_v \frac{E_c - \varepsilon_\Psi }{E_p} + \alpha_c \alpha_v
 \frac{\hbar^2}{2 m_0 E_p} \left( k_1^2(\varepsilon_\Phi)+
 k_1^2(\varepsilon_\Psi \right)  \right) \right. \, -  \nonumber \\
 &-& \left. \Psi_{v}^{*} \Phi_{c}  \left( 1 + \alpha_c
 \frac{\varepsilon_\Psi - E_v}{E_p} + \alpha_v \frac{E_c -
 \varepsilon_\Phi }{E_p} + \alpha_c \alpha_v \frac{\hbar^2}{2m_0 E_p}
 \left( k_1^2(\varepsilon_\Phi)+ k_1^2(\varepsilon_\Psi )  \right)
 \right) \right] = Const \, . \label{Jnon1}
\end{eqnarray}
Here $\varepsilon_\Psi$ and $\varepsilon_\Phi$ are the energies of
the states described by the two component envelope functions
${\Psi}_1(k_1) \equiv (\Psi_c=\Psi_c(k_1),\Psi_v=\Psi_v(k_1))$ and
${\Phi}_1(k_1)\equiv (\Phi_c=\Phi_c(k_1),\Phi_v=\Phi_v(k_1))$
respectively. The two components of the normal velocities $V_{\Psi
\, c}$ ($V_{\Phi \, c}$) and  $V_{\Psi \, v}$ ($V_{\Phi \, v}$) are
expressed through  the wave function components $\Psi_c$ ($\Phi_c$)
and $\Psi_v$($\Phi_v$). Since the envelope functions $\Psi$ and
$\Phi$ are now truncated, the normal velocity and envelope function
components  are no longer linearly independent. Therefore, the Eq.
(\ref{Jnon1}) can not  be fulfilled if $\alpha_{c} \alpha_{v} \ne
0$. (For the case $\alpha_{c} \alpha_{v} = 0$ the spurious solution
$k_2$ does not arise,  and ${\Psi}_1$ is then the full solution of
the Hamiltonian (\ref{KanedH})).  However we shall see
 below that we can impose an alternative set of the state
independent BC that retain the continuity of the normal flux density
$J_{\tau, 1}^{\Psi\, \Phi}$ for truncated wave functions within an
accuracy consistent with  the MEMA.  We consider a  $\tilde J_{\tau,
1}$ that differs from $J_{\tau, 1}$ of Eq. (\ref{Jnon1})  only by
the   terms $\tilde \gamma^2\sim \alpha_c\alpha_v$
 which are   small:  $\tilde \gamma^2
= \alpha_c \alpha_v (\varepsilon - E_c) (\varepsilon -
E_v)/E_p^2\approx (k_1/k_2)^2$.   A typical value of the  momentum
$k_1$ in a heterostructure with  size $L$ is $k_1\sim 1/L$.
Substituting the value  of $k_2$ from Eq.(22) into $\tilde \gamma$
we find that $\tilde \gamma^2\sim (a_0/L)^2\ll 1$. We can write then
the  approximate condition
\begin{eqnarray}
& & \tilde J_{\tau , 1}^{\Psi \, \Phi}(z) = \frac{i \, P}{m_0}  \left[
\Psi^{*}_{c}  \Phi_{v} \left( 1 + \alpha_c \frac{\varepsilon_\Phi
- E_v}{E_p}  \right) \left( 1 + \alpha_v \frac{E_c -
\varepsilon_\Psi }{E_p}  \right) \right. \, -  \nonumber \\ &-&
\left. \Psi_{v}^{*} \Phi_{c}  \left( 1 + \alpha_c
\frac{\varepsilon_\Psi - E_v}{E_p}\right) \left(  1 + \alpha_v
\frac{E_c - \varepsilon_\Phi }{E_p} \right) \right] =  Const \, ,
 \label{Jnon2}
\end{eqnarray}
which is  fulfilled exactly if
the  envelope functions $\Psi$ and $\Phi$ satisfy the
 truncated set of  BC:
\begin{equation}
\left ( \begin{array}{c} \Psi^{r}_c \left( 1 + \alpha_v^{\it r}
\frac{E_c^{\it r} - \varepsilon}{E_p^{\it r}} \right) \\ P^{\it r}
\Psi^{\it r}_v \left( 1 + \alpha_c^{\it r} \frac{\varepsilon-
 E_v^{\it r}}{E_p^{\it r}} \right)
 \end{array} \right) = \tilde T_{tr} \left(
\begin{array}{c}
\Psi^{l}_c \left( 1 + \alpha_v^l \frac{E_c^l - \varepsilon}{E_p^l}
\right)\\ P^l \Psi^{l}_v \left( 1 + \alpha_c^l \frac{\varepsilon-
E_v^l}{E_p^l} \right)
\end{array} \right) \,
,
 \label{truncBC}
\end{equation}
where  the components of the  $2 \times 2$ transfer matrix  $\tilde
T_{tr}$ are real and state independent and $det[\tilde T_{tr}]
= 1$. One can see that the connection between the left and right
hand side components of $\Psi_c$ and $\Psi_v$  has now energy
dependent coefficients. This is a consequence of the  linear
dependence  of the components of the truncated wave function and its
normal velocity on the same side of the interface. However, the new
energy independent transfer matrix $\tilde T_{tr}$ is still
characteristic of the interface only.  The truncated general BC of
Eq. (\ref{truncBC}) no longer allow the simultaneous continuity of
the both envelope function components $\Psi_c(k_1)$ and
$\Psi_v(k_1)$ at the interface.  For the case where the  off
diagonal matrix elements of $\tilde T_{tr}$ are negligibly small,
$\tilde t_{12}\approx\tilde t_{21}\approx 0$, the general BC reduce
to:
\begin{equation}
P^{\alpha} \left( 1 + \alpha_v \frac{E_c - \varepsilon}{E_p}
\right) \Psi_c = Const \, , \qquad  P^{1-\alpha}  \left( 1 +
\alpha_c \frac{\varepsilon-  E_v}{E_p} \right) \Psi_v = Const \, ,
\label{Kanedlra}
\end{equation}
which is similar to the BC of Eq. (\ref{flra}) for heterostructures
with parabolic bands.

Equations (\ref{truncBC}, \ref{Kanedlra}) are general BC for the
truncated smooth wave functions in semiconductor heterostructures
with non-zero $\alpha_c \alpha_v $. These BC do not depend on the sign
of $\alpha_c \alpha_v$ (i. e., whether the spurious
solution is evanescent or propagating). They  are also valid
for any Kane model with $\alpha_c \alpha_v = 0$, in which
 Eqs. (\ref{selffluxdk},\ref{Jnon1}) are satisfied
exactly. The boundary condition Eq.
(\ref{Kanedlra})  was first discussed in Ref. \onlinecite{ivchenko}
 for the "pure" Kane model with
$\alpha_c=\alpha_v=0$;
also the model with $\alpha_{c}=0$, $\alpha_{v} \ne 0$
was considered in Ref. \onlinecite{foreman}.

To generalize this procedure  and derive general BC for truncated
slowly varying envelope functions for heterostructures described by
an $N\times N$ Hamiltonian having $m$ spurious solutions, it is
important to  trace from where these solutions arise. Examining Eqs.
(17,\ref{kroots}) (and also see Ref. \onlinecite{gelmont} which
describes spurious solutions in the $8 \times 8$ model) we find that
the spurious solutions $|k_i|$ ($i=N-m+1 \, , \dots N$) of the
dispersion equations for the Hamiltonian of Eq. (\ref{Nham}) are
proportional to the large components of the matrices $\hat {\bf B}$
and inversely proportional to the square root of the product of the
tensor $D_{\mu \nu}$ conduction and valence band components. For the
sake of definiteness \cite{small}, let us assume that the matrices
$\hat {\bf B}$ contain    the submatrices $\hat {\bf  B}^{'}= \hat
{\bf  B}^{\gamma}/\gamma$ of rank $2m$, where $\gamma \sim a_0/L$ is
a dimensionless small parameter responsible for the  spurious
solutions.  For  example, in the two band model considered above the
normal component $ B_\tau^{\gamma}= \vec{\tau} \cdot \hat {\bf
B}^{\gamma}$  is:
 \begin{equation}
{\hat B_\tau^\gamma}=   \frac{i\,\hbar}{m_0 L}  \left(
\begin{array}{cc}
0&1 \\
- 1 &0
\end{array} \right) \, , \quad \gamma=\frac{\hbar}{P L}\sim \frac{a_0}{L} \, .
\label{velgendH}
\end{equation}
We can write the velocity operator of Eq. (\ref{efavelocity}) as
\begin{equation}
\hat {\bf V} = \frac{1}{\gamma} \hat {\bf B}^{\gamma} + \hat {\bf
V}_0 \, , \label{velgen}
\end{equation}
and write the truncated slowly varying envelope function  as the
superposition of two linearly independent  orthogonal wave functions
$\tilde {\Psi}={\Psi}_1 + {\Psi}_2$, such that the $2m$ component
function ${\Psi}_1$ and $n=N-2m$ component function ${\Psi}_2$
satisfy the condition
\begin{equation}
B_\tau^{\gamma} {\Psi}_1 \ne 0 \, , \qquad  B_\tau^{\gamma}
{\Psi}_2 = 0 \, . \qquad
\end{equation}
Furthermore, we can write the function $\vec{\tau} \cdot \hat {\bf
V}_0 \tilde {\Psi}=V_1+V_2$ as superposition of a $2m$ component
function ${V}_1$ and an $n$ component function ${V}_2$ belonging to the same
subspaces as the functions ${ \Psi}_1$ and ${\Psi}_2$, respectively,
i.e., $({\Psi}_1,{V}_2) = ({\Psi}_2,{V}_1)=0$. Note that $\Psi_2$
is absent ($n=0$) in the
two  band model; $\Psi_1$ is the total
slowly varying envelope function and $V_{1}=\hat V_0 \Psi_1$ ($\hat
V_0$ is the diagonal part of the velocity operator in Eq.
(\ref{velocitydH})). However, the functions ${V}_1$ and ${V}_2$ do
not coincide with $\vec{\tau} \cdot \hat {\bf V}_0
\Psi_1$ and $\vec{\tau} \cdot \hat {\bf V}_0 \Psi_2$,  respectively,
 for the general
case $n \ne 0$. The flux density continuity of Eq.
(\ref{generalflux}) can now be written  as:
\begin{eqnarray}
J_\tau^{\alpha \, \beta}&=& J_{{\tau},1}^{\alpha \, \beta}+
J_{{\tau},2}^{\alpha \, \beta}= Const \, , \label{jconst} \\
J_{{\tau},1}^{\alpha \, \beta}&=& \frac{1}{\gamma} \left(
{\Psi}_{1\, \alpha},  B_\tau^{\gamma} {\Psi}_{1 \, \beta} \right)
+ \frac{1}{2}\left[ \left( {\Psi}_{1 \, \alpha}, {V}_{1 \, \beta}
\right) + \left( {V}_{1\, \alpha}, {\Psi}_{1\, \beta} \right)
\right] \, , \label{j1} \\
J_{{\tau},2}^{\alpha \, \beta}
&=&\frac{1}{2}\left[ \left( {\Psi}_{2 \, \alpha}, {V}_{2 \, \beta}
\right) + \left(  {V}_{2\, \alpha}, {\Psi}_{2\, \beta} \right)
\right]~.
\label{generalfluxtrunc}
\end{eqnarray}
Neglecting terms of second order in $\gamma$,
Eq. (\ref{j1}) can be expressed
\begin{eqnarray}
\tilde J_{{\tau},1}^{\alpha \, \beta}= \frac{1}{\gamma} \left(
{\tilde \Psi}_{1\, \alpha},  B_\tau^{\gamma} {\tilde \Psi}_{1 \,
\beta} \right) \, , \qquad  \tilde \Psi_1= {\Psi}_{1} +
\frac{\gamma}{2}  B_\tau^{\gamma\, -1} V_{1} \, ,
\label{generalfluxtrunc1}
\end{eqnarray}
 where $V_{1}$   is, to this order,  included  in ${\tilde
\Psi}_{1}$. Equations (\ref{j1}) and  Eq. (\ref{generalfluxtrunc1})
are the multiband extension of  Eqs. (\ref{Jnon1}) and (\ref{Jnon2})
obtained  for the two band model.  The vector $\{{\tilde \Psi}_1,
{\Psi}_2, V_{2}\}$ has in total $2m+2n=2(N-m)$ independent
components and the general boundary condition satisfying
$J_{{\tau}}^{\alpha \, \beta} \approx \tilde J_{{\tau}}^{\alpha \,
\beta} = \tilde J_{{\tau},1}^{\alpha \, \beta}+ J_{{\tau},2}^{\alpha
\, \beta}= const$ can be written as:
\begin{equation}
\left( \begin{array}{c}
{\tilde \Psi}_1^{+} \\{\Psi}_2^{+} \\ i {V_{2}^{+}}
\end{array} \right) = T_{2(N-m)} \left(
\begin{array}{c}
{\tilde \Psi}_1^{-} \\{\Psi}_2^{-} \\ i {V_{2}^{-}}
\end{array} \right)
, \qquad  T_{2(N-m)}= \left( \begin{array}{c c}
T_{2m} & T_{12} \\ T_{21} & {T}_{2n} \end{array} \right)
 \, .
\label{matrixgentrunc}
\end{equation}
 We shall neglect the off diagonal  matrices, i.e., set
$T_{12}=T_{21}=0$, (thus satisfying the stronger condition that both
$\tilde J_1^{\alpha \beta}=const$ and $J_2^{\alpha \beta}=const$)
and obtain the following restrictions on the elements of the energy
independent  transfer matrices $T_{2m}$ and $T_{2n}$:
\begin{equation}
T_{2m}^\dagger \left. \frac{1}{\gamma} B^{\gamma}_\tau
\right|_{{\bf S}_j+0} T_{2m} = \left. \frac{1}{\gamma}
B^{\gamma}_\tau \right|_{{\bf S}_j-0} \, , \qquad
{T}^{\dagger}_{2n} \left(  \begin{array}{c c} 0 &  I_{2n} \\ -
I_{2n} & 0
\end{array} \right) {T}_{2n} = \left(  \begin{array}{c c}
0 & I_{2n} \\ - I_{2n} & 0
\end{array} \right) \, .
\label{matrixgen12}
\end{equation}

Equations (\ref{matrixgentrunc}) and (\ref{matrixgen12}) represent
$2(N-m)$ boundary condition equation for the truncated envelope wave
function. These  general BC do not allow the  continuity of
all envelope function components at the interface if
$B_{\tau}^\gamma/\gamma$ is not continues. While the elements of
$T_{2m}$ and $T_{2n}$  are state independent, the  function $\tilde
\Psi_1$ may be related to $\Psi_1$ by the  energy dependent expression.

  This procedure can be directly applied  to
heterostructures described by the $8 \times 8$ Pidgeon and Brown
Hamiltonian \cite{PB}. The bulk energy dispersion equations for this
Hamiltonian has 8  solutions, $k^2_i$, (taking spin degeneracy into
account), for each energy $\varepsilon$; two of these are spurious,
and proportional to the large interband Kane matrix element $P$
(see, for example, Ref. \onlinecite{gelmont}). In the case of zero
spin--orbit interaction, $\Psi_1$ must include  coupled electron and
 light hole band envelope
functions, while
$\Psi_2$ includes the heavy hole band envelope functions
only. For planar heterostructures Eqs. (\ref{matrixgentrunc},
\ref{matrixgen12}) take  into account the mixing of light and heavy
holes  for the states with finite in-plane momentum.  However they
neglect the effect of the "low interface symmetry" \cite{ivkamros}
that mixes light and heavy hole plane waves normally incident on
the  interface (because we assume that $T_{12}=T_{21}=0$).  The
detailed application of our procedure to the $8 \times 8$ models
for planar and spherical heterostructures will be presented
elsewhere.

\section{One dimensional quantum well  with infinite potential barriers.}
\label{calc}

Let us consider an effect of the general BC  on the energy spectrum
of a one dimensional quantum well having infinite potential
barriers.  The conventional BC for structures with an
impenetrable barrier require  vanishing of the envelope wave
function at the barrier surface, but this requirement has
 never been justified (see
for example Ref. \onlinecite{brag99}).  In general,
 the self--adjointness
of the MEMA Hamiltonian  requires vanishing of the normal
flux density at the boundaries:
 \begin{equation}
J_\tau^{\alpha \beta} ({z= \pm L \mp 0})  = 0 \, ,
\label{infflux}
\end{equation}
where $2L$ is the thickness of the quantum well. If we assume that
the two interfaces are completely identical, the quantum well
possesses reflection symmetry about $z=0$ and all solutions of the
Schr\"{o}dinger equation are characterized by their parity according
to whether  their components are  even or odd under reflection.
 However, the BC are  local
characteristics of each interface.  Therefore, the  Eq. (\ref{infflux})
must be fulfilled  at each
interface independent of the symmetry of the functions
 $\Psi_\alpha$ and $\Psi_\beta$.

Now we consider the effect of the general BC  for the case of the two
band Kane model.  In order to satisfy Eq. (\ref{infflux}) in a
quantum well with symmetric interfaces, one  can write the general
BC for the conduction $\Psi_c$ and valence band $\Psi_v$ components
of the envelope wave  function (which do not contain spurious
solutions) as
\begin{eqnarray}
\Psi_c (\pm L \mp 0) \left( 1 - \alpha_v
\frac{\varepsilon-E_c}{E_p} \right) = \mp \Theta \, \Psi_v (\pm L \mp
0) \left( 1 + \alpha_c \frac{\varepsilon - E_v }{E_p} \right) ,
 \label{truncBCinf}
\end{eqnarray}
 where $\Theta$ is a real number and the sign difference is due to the
opposite parity  of the conduction and valence band components of
the envelope functions.   The surface parameter $\Theta$  does not
depend on the energy or symmetry of either state. Equation
(\ref{truncBCinf}) can be derived directly from the general BC of
Eq. (\ref{truncBC}) obtained in Sec. \ref{spur} for the case of a
finite potential barrier. To do this one has to assume that $\Psi
\equiv 0$ outside the quantum well and use the state independent
transfer matrices $\tilde T_{tr}^+$, for $z=L-0$, and $\tilde
T_{tr}^-$, for  $z=-L+0$ interfaces, which differ only by the sign
(opposite) of the  offdiagonal matrix elements and satisfy  the
condition $\det[\tilde T_{tr}^+]=\det[\tilde T_{tr}^-]=0$.

 Neglecting
$\gamma^2 \sim (a_0/L)^2$ terms, Eq. (\ref{truncBCinf}) can be
written as two separate equations for the conduction and valence
band components, respectively
\begin{eqnarray}
\Psi_c (\pm L \mp 0 ) = \pm T_c a_0 \frac{m_0}{m_c(E_{e})}
\Psi_c^{'}( \pm L \mp 0) \, ,  \qquad  T_c a_0= \Theta \,
\frac{\hbar}{2P} \, ,\\  \Psi_v (\pm L \mp 0 ) = \pm T_v a_0
\frac{m_0}{m_v(E_{h})} \Psi_v^{'}( \pm L \mp 0) \, ,  \qquad T_v
a_0= \frac{1}{\Theta} \, \frac{\hbar}{2P} \, .
\label{infbc}
\end{eqnarray}
Here the  energy--dependent effective masses in the conduction and
valence bands are given by
\begin{eqnarray}
\frac{m_0}{m_c(E_{e})}=
\alpha_c + \frac{E_p}{E_{e} + E_g} \,,  \quad
\frac{m_0}{m_v(E_{h})}=  \alpha_v + \frac{E_p}{E_{h}
+ E_g} \, ,
\label{massbc}
\end{eqnarray}
where the electron, $E_{e}$,  and hole $E_{h}$ energies are measured
from the bottom of conduction and the top of the valence bands,
respectively: $E_{e}=\varepsilon - E_c$ and $E_{h}=E_v -
\varepsilon$. The even, (+), and odd, (-),  solutions to Eqs.
(38,39) can be written
\begin{eqnarray} \Psi_{c(v)}^+(z) = A_{c(v)} \,
\cos(\phi_{c(v)}^+ z) \, , \qquad  \Psi_{c(v)}^-(z) = A_{c(v)}
\, \sin(\phi_{c(v)}^- z) \, ,
\end{eqnarray}
where $A_{c,v}$ are the  normalization constants.  We can derive
equations for the energies of the even  and odd electron and hole
quantum size levels:
\begin{eqnarray}
E_{e}^{\pm}= \frac{\hbar^2 (\phi_{c}^{\pm})^2}{2 m_0 L^2}
\frac{m_0}{m_c(E_{e}^\pm)}  \left( 1- \alpha_v
\frac{E_{e}^\pm}{E_p} \right) \, , \quad  E_{h}^{\pm}=
\frac{\hbar^2 (\phi_{v}^{\pm})^{2}}{2 m_0 L^2}
\frac{m_0}{m_v(E_{h}^\pm)} \left( 1- \alpha_c
\frac{E_{h}^\pm}{E_p} \right) \, , \label{qsl}
 \end{eqnarray}
where $\phi_{c}^\pm$ and $\phi_{v}^\pm$ are the solutions of the
equations:
\begin{equation}
\phi_{c}^\pm [\tan (\phi_{c}^\pm)]^{\pm 1} =
\mp \frac{m_{c}(E_{e}^\pm)}{m_0}
\frac{L}{T_c a_0}~,~~  \phi_{v}^\pm [\tan (\phi_{v}^\pm)]^{\pm 1} =
\mp \frac{m_v(E_{h}^\pm)}{m_0} \frac{L}{T_v a_0} \, .
\label{rootbc}
\end{equation}

\begin{figure}
\setlength{\unitlength}{1cm}
\begin{center}
\begin{picture}(12,12)
\epsfxsize10cm \put(0,0){\epsffile{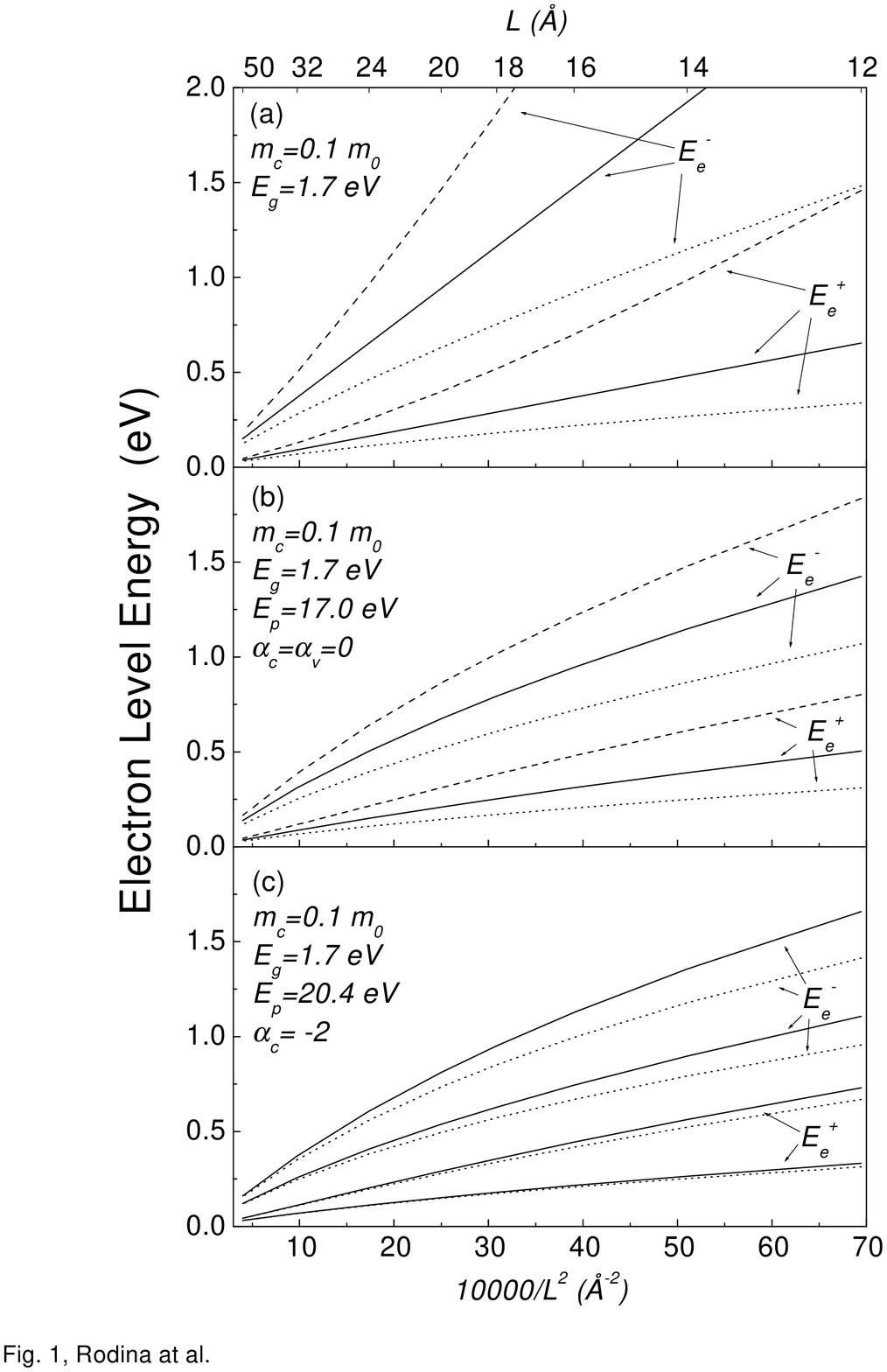}}
\end{picture}
\end{center}
\caption{The size dependence of the two lowest, even ($E_e^+$) and
odd ($E_e^-$), electron  quantum size levels in a well with infinite
potential barriers calculated as a function of the surface parameter
$Ta_0$: (a)  in the parabolic EMA model; (b) in the pure Kane model
($\alpha_c=\alpha_v=0$); (c) in the full two band model ($\alpha_c
\alpha_v \ne 0$). In (a) and (b), the solid, dashed and dotted
curves  correspond to $Ta_0=0$, $Ta_0=0.5$ and $Ta_0=-0.5$\,\AA,
respectively. In (c) the  solid and dotted curves  are calculated
with $\alpha_v=-2$ and $\alpha_v=2$, respectively, upper curves for
each level correspond to $Ta_0=0.43$\,\AA\ and lower curves to
$Ta_0=-0.43$\,\AA. Other parameters used in calculations are shown
in the Figure. } \label{levL}
\end{figure}

Figure \ref{levL} shows the dependence of the two lowest quantum
size  electron levels, $E_e^+$ and $E_e^-$, on the well width $L$ as
a function of  of the surface parameter $T_c a_0$. In these
calculations we use for $m_c$ the electron effective mass at the
bottom of the conduction band  $m_c=0.1 m_0$,  and a  band gap
energy $E_g=1.7$\,eV which are close to the parameters of CdSe and
CdTe. We compare the effect of the general BC  on the quantum size
levels in the parabolic EMA  (Fig.1a), in the "pure" Kane model with
$\alpha_c=\alpha_v=0$ (Fig. 1b), and in the "full" two band model
with $\alpha_c \alpha_v \ne 0$ (Fig.1c).

Equations (\ref{qsl},\ref{rootbc}) describe the energy of the
quantum size levels  for coupled conduction and valence bands.
However, the conduction and valence band energy spectra can be
considered as being independent when the energies of the electron or
hole levels  are much less than the energy gap, $|E_{e}|,|E_{h}| \ll
E_g, E_p$. This limit case is realized in thick quantum  wells and
is described by the equations for the simple parabolic bands that
one obtains by  neglecting the energy dependence of the effective
masses  and the $E_{e,h}/E_p$  terms in Eqs.
(\ref{qsl},\ref{rootbc}). For this case the surface parameters $T_c
a_0$ and $T_v a_0$ can be chosen independently, and the conventional
BC $\Psi_c(\pm L)=0$ are realized  for $T_c a_0 = 0$.  Fig.
\ref{levL}(a) shows the effect of {\bf $T_c a_0  \ne 0$}   on the
electron quantum size levels in the parabolic EMA.  One sees that
positive and negative values of $T_c a_0$ shift the quantum size
levels energy   up and down, respectively, from those obtained using
conventional BC. The effect is negligible when $|T_c a_0|/L \ll
\phi_c^\pm\, m_0/m_c$ but become noticeable in narrow wells and is
greater for higher energy levels.

Fig. \ref{levL}(b) shows the effect of the general BC on the
electron quantum size levels in the pure Kane model. One sees that
the effect of $T_ca_0 \ne 0$ is very similar to that in the
parabolic EMA. The size dependence differs only in the nature of
the nonlinear dependence on $1/L^2$. This is because the Kane model
takes the nonparabolicity of the conduction band into account.

The size dependence of the electron levels calculated with $T_c
a_0=0$ is shown in Fig. \ref{levL}(b) only for comparison. The
conventional BC do not hold in the Kane model, because
the surface parameters for coupled conduction and valence bands
are related  by $(T_c a_0)(T_v a_0) = \hbar^2/2E_p m_0$.   Choosing
$T_c a_0=0$  for determining the electron energy levels  corresponds
to choosing $T_v a_0 \rightarrow \infty$, which does not describe
the hole energy levels. The condition  $|T_c a_0|=|T_v
a_0|=a^*$ realized for $|\Theta|=1$, describes a symmetric
(relative to the to the center of the band gap)  structure of the
electron and hole  quantum size levels in semiconductors with
$\alpha_c=\alpha_v$. Although these "symmetric semiconductor
structures" do not exist in nature, the parameter $a^*=
\hbar/2P\approx 0.45 \pm 0.06$\,\AA\ gives a reasonable value of
$|T_c a_0|$ and $|T_v a_0|$ in real semiconductor structures. If
symmetric BC hold
\begin{eqnarray}
{\phi_{c}^{\pm}}\, \frac{m_0}{m_{c}(E_{e})} \frac{a^*}{L}
\ll 1 \, ,  \qquad {\phi_{v}^{\pm}}\, \frac{m_0}{m_{v}(E_{h})}
\frac{a^*}{L}  \ll 1 \,   ,
\end{eqnarray}
and the solutions of Eq. \ref{rootbc} for the lowest electron and hole
quantum size levels, $\phi_{c,v}^{\pm}$,  are close to those given
by $T_{c,v} a_0 = 0$.

Figure \ref{levL} (c) shows the size dependence of the electron
levels calculated in the full two  band model with  the symmetric
surface parameter   $|Ta_0|=a^*=0.43$\,\AA. One can see that the term
linear in $\alpha_v$ in Eq. (\ref{qsl}) for the electron energy
levels becomes important when they comparable with the band
gap energy. On the other hand changing the sign of $\alpha_v$ leads
only  to small changes of the level energy.

\begin{figure}
\setlength{\unitlength}{1cm}
\begin{center}
\begin{picture}(12,12)
\epsfxsize10cm \put(0,0){\epsffile{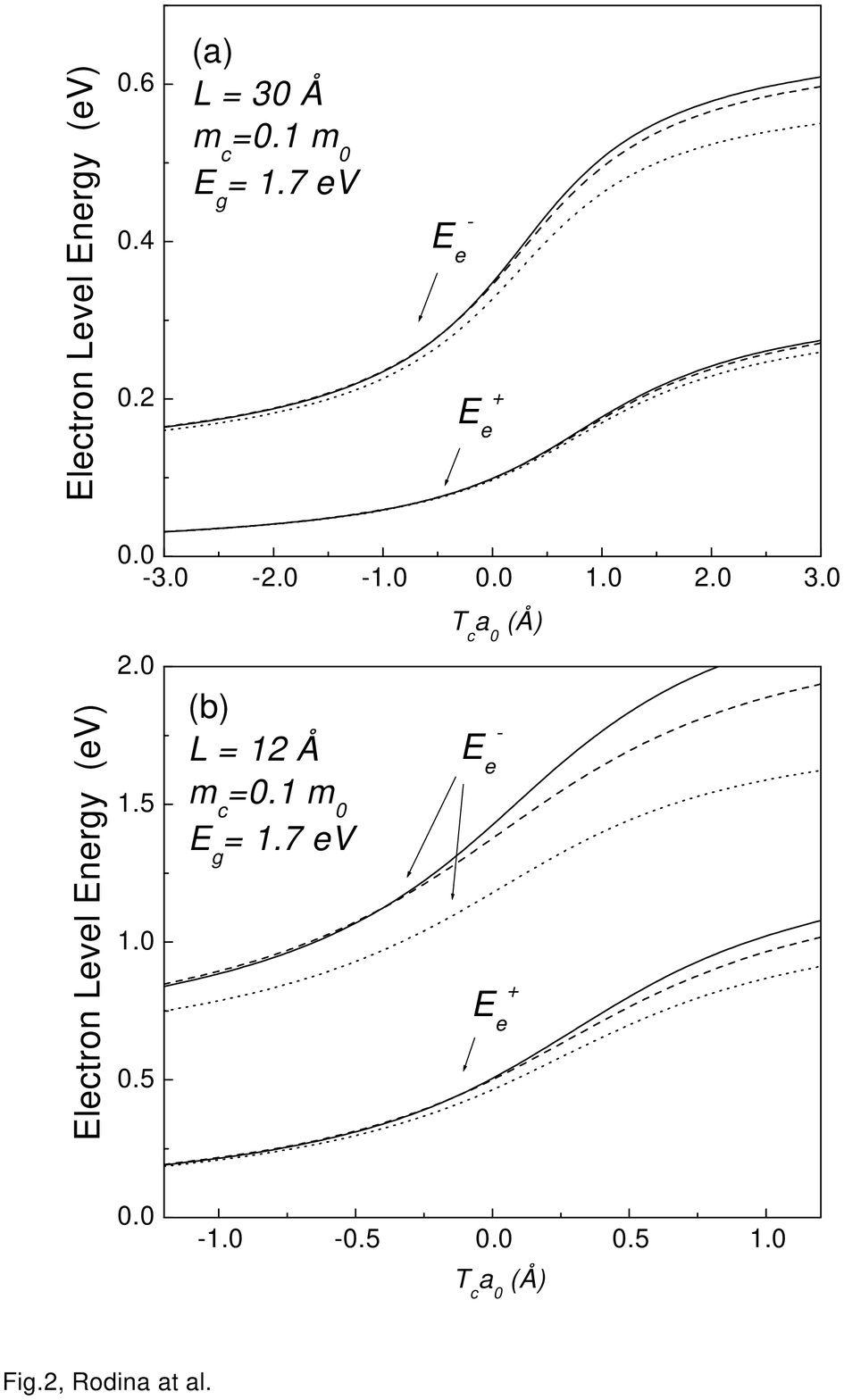}}
\end{picture}
\end{center}
\caption{The dependence of the two lowest , even ($E_e^+$) and odd
($E_e^-$), electron  quantum size levels on the surface parameter,
$T_c a_0$,  calculated in a well of width $2L$ surrounded by
infinite potential barriers: (a) $L=30$\,\AA\ and (b) $L=12$\,\AA.
The solid line is for  the pure Kane model with
$\alpha_c=\alpha_v=0$ and $E_p=17.0$\,eV, the dashed line is for the
symmetric two band model with $\alpha_c=\alpha_v=-2$ and
$E_p=20.4$\,eV, and the dotted line for the asymmetric two band
model with $\alpha_c=-2$, $\alpha_v= 2$ and $E_p=20.4$\,eV. Other
parameters used in the calculations are shown in the Figure.}
\label{levT}
\end{figure}

The dependence of the lowest electron  quantum size levels on the
surface parameter $T_c a_0$ for the Kane and full two band models
for wells  with $L=30$\,\AA\ and $L=12$\,\AA, respectively, is
shown in  Fig. \ref{levT}. The surface parameter is varied from
$-3.0$\,\AA\ to $3.0$\,\AA\ for  $L=30$\,\AA,  and from $-1.2$\,\AA\
to $1.2$\,\AA\ for  $L=12$\,\AA, respectively, so that  $|T_c a_0|
\le L m_c/m_0$ is fulfilled. One can see that varying the surface
parameter  in this range produces  a monotonic change of the first
even and odd electron levels. The difference between models with
different $\alpha_c$ and $\alpha_v$ (for the same $m_c$) is   small
for $L=30$\,\AA, but can be important for $L=12$\,\AA.

It is interesting to note that for positive values of the  surface
parameter $\Theta$ ($T_c a_0>0$, $T_v a_0>0$),  equations
(\ref{qsl},\ref{rootbc}) may have even and odd solutions with  an
energy in the forbidden  gap: $E_{e} <0$, $E_{h} <0$. In wide wells
the energy of these gap levels do not depend on their symmetry,
$E_{(e,v),S}^+=E_{(e,v),S}^-=E_{(e,v),S}$, and can be found from
\begin{eqnarray}
E_{e,S}  = - \frac{\hbar^2}{2m_0} \frac{m_c(E_{e,S})}{m_0}
\frac{1}{(T_c a_0)^2}\cdot \left( 1- \alpha_v \frac{E_{e,S}}{E_p}
\right)\, , \label{gape}\\ E_{h,S}  = - \frac{\hbar^2}{2m_0}
\frac{m_v(E_{h,S})}{m_0} \frac{1}{(T_v a_0)^2}\cdot \left( 1-
\alpha_c \frac{E_{h,S}}{E_p} \right)\, . \label{gaph}
\end{eqnarray}
In wide "symmetric wells" these gap states have the same energy
$E_{e,S}=E_{h,S}=-E_g/2$, and are localized within a layer of
thickness $a_s=2a^* E_p/E_g$ near the surface.   These solutions
contain no  contributions from the unphysical spurious solutions,
and thus  are not artifacts of the ${\bf k} \cdot {\bf p}$ model.

\section{Discussion and comparison with other approaches}
\label{compar}

The occurrence of discontinuities in the   envelope wave functions
at the heterointerface is one of the most important consequence of
general BC.  These discontinuities have a strong effect on the
mathematical procedures often used for the calculation of various
physical properties of heterostructures having finite potential
barriers. In these procedures, the Sch\"{o}dinger equation  is
integrated   across the interface or a Fourier transformation is
performed using the piece--wise spatially determined material
parameters   with the help of  generalized step functions. However,
the integration of terms such as products of $\frac{1}{m} \hat
k_z^2$ or $P \hat k_z$ with envelope wave functions, $\Psi$, which
are discontinuous  across an interface   may lead to  mathematical
uncertainties: integration of the product of a step function and a
$\delta$--function is not well defined. To resolve this problem, a
non-unique\cite{morrow,winkler,balian,pistol}   symmetrized form of
the kinetic energy operator $\hat H_k$ is used, and the BC  for the
envelope function $\Psi$ are obtained by requiring that  $\hat H_k
\Psi$ be integrable across the interface \cite{morrow,baraff}.

For example, in order to obtain the general BC of Eq.(4) for
parabolic bands, symmetrized  kinetic  energy operator of
the form   \cite{morrow}
 \begin{equation}
\hat H^{\alpha}_K = -\frac{\hbar^2}{2} m^\alpha
\frac{d}{dz}\frac{1}{m^{1+2\alpha}}\frac{d}{dz} m^\alpha \,
\label{efa}
\end{equation}
is used. The case $\alpha=0$ then leads to  the conventional BC of Eq.
(\ref{flr}). The same symmetrized form, $\hat k_z \frac{1}{m} \hat
k_z$, is usually used for the diagonal terms of the multiband
kinetic energy operator \cite{baraff,winkler}. To derive the BC of
Eq. (\ref{Kanedlra}) for the "pure" two band Kane model
($\alpha_c=\alpha_v=0$) that allow to integrate Schr\"odinger
equation across the interface,  one
writes  the kinetic energy operator as:
\begin{equation}
\hat H_K^\alpha= \frac{i \hbar}{m_0} \left(
\begin{array}{cc}
0 &     P^\alpha \hat k_z P^{1-\alpha} \\ -
 P^{1-\alpha} \hat k_z P^{\alpha} & 0 \end{array}
\right) \, .
\label{KaneHKa}
\end{equation}
Here, now, there is no value of $\alpha$ that gives the symmetrized
form $1/2(\hat k_z P + P \hat k_z )$, that is usually used for the
off diagonal terms (linear in $k$) of the multiband
Hamiltonian.\cite{baraff,winkler} An asymmetric ordering,
corresponding to $\alpha=1$  was suggested in Ref.
\onlinecite{foreman} for the model with $\alpha_c=0$.  The $\alpha$
dependence of Eqs.  (\ref{efa}) and (\ref{KaneHKa}) clearly
demonstrates that  it is  the BC that  determine the  integration
across the abrupt heterointerface  and not the other way round.

Fourier  transforming the Schr\"{o}dinger equation has been
suggested by Winkler and R\"{o}ssler \cite{winkler} as an
alternative approach  to the MEMA problem for heterostructures with
finite potential barriers. Then one does not deal explicitly  with
BC  when solving the resulting integral equations for the  momentum
space envelope function, and  avoids unphysical spurious solutions
by  restricting the range of integration to $|k| \ll 2\pi /a_0$.
However, the resulting form of the momentum space MEMA Hamiltonian
depends on the particular symmetrization procedure chosen for the
kinetic energy operator in configuration space \cite{winkler}, and,
therefore, is again determined by the BC imposed on the envelope
function.  To illustrate this, we derive the explicit form of the
momentum space Schr\"{o}dinger equation for the one band EMA and the
two band "pure" Kane model, using general BC. Fourier integration of
the EMA Schr\"{o}dinger equation with the kinetic energy operator of
Eq. (\ref{efa}) leads to the following  integral equation for the
momentum space envelope function $F(k)$:
 \begin{eqnarray}
\frac{\hbar^2}{2} \int_{-\infty}^{\infty}q^{'} k^{'}
M_{-\alpha}(k-q^{'})M_{1+2\alpha} (q^{'}-k^{'})
M_{-\alpha}(k^{'}-q) F(q)  dq^{'} dk^{'}\, dq +  \nonumber \\ +
\int_{-\infty}^{\infty} F(q) V_c(k-q)dq = \varepsilon F(k) \, ,
\label{furalpha}
\end{eqnarray}
where $F(k)$ satisfies the normalization condition $
\int_{-\infty}^{\infty} | F(k) |^2 dk = 1$, and
\begin{eqnarray}
V_c(k)= \frac{1}{2\pi} \int_{-\infty}^{\infty} E_c(z) \exp(-ikz) dz \,
, \qquad
M_\nu(k)= \frac{1}{2\pi} \int_{-\infty}^{\infty} \frac{1}{m^\nu(z)}
\exp(-ikz) dz \, .
\end{eqnarray}
Here $E_c(z)$ and $m(z)$ are the energy of the bottom of the
conduction band and the electron effective mass in each
region of the one-dimensional heterostructure considered.  The
configuration space envelope function, in the $j$--th region of the
heterostructure, is now given by
\begin{eqnarray}
f^j(z) =  m_j^{-\alpha} \cdot \frac{1}{\sqrt{2\pi}}
\int_{-\infty}^{\infty} F(q) M_{-\alpha}(k-q) \exp(ikz) dk dq \, ,
\end{eqnarray}
and satisfies the general BC of Eq. (\ref{flra}). For the "pure" Kane
model, Fourier transforming the Schr\"{o}dinger equation, using the
kinetic energy operator of Eq. (\ref{KaneHKa}),  leads to the
following coupled integral equations for the two component momentum
space function $\{F_c(k),F_v(k)\}$ :
\begin{eqnarray}
\int_{-\infty}^{\infty} F_c(q) V_c (k-q) dq  + i \frac{\hbar}{m_0}
\int_{-\infty}^{\infty} q' G_{\alpha} (k-q') G_{1-\alpha} (q' -q)
F_v(q) dq dq' = \varepsilon F_c(k) \, ,\\ - i \frac{\hbar}{m_0}
\int_{-\infty}^{\infty} q' G_{1-\alpha} (k-q') G_{\alpha} (q'-q)
F_c (q) dq dq' + \int_{-\infty}^{\infty} F_v (q) V_v (k-q) dq =
\varepsilon F_v(k) \, , \label{KaneFur}
\end{eqnarray}
where $\int_{-\infty}^{\infty}
\left( | F_c(k) |^2 + | F_v(k) |^2 \right) dk  = 1$, and
\begin{eqnarray}
V_{c,v}(k)= \frac{1}{2\pi}\int_{-\infty}^{\infty}  E_{c,v}(z)
\exp(-ikz) dz \, , \quad G_\nu(k)= \frac{1}{2\pi}
\int_{-\infty}^{\infty} P^\nu(z) \exp(-ikz) dz \, .
\end{eqnarray}
Here $E_v(z)$ and $P(z)$ are the energy of the top of the valence
band and the Kane matrix element  in each region, respectively. The
configuration space envelope function $\Psi=\{\Psi_c,\Psi_v\}$ in the
$j$--th region of the  one-dimensional heterostructure is now
given by
\begin{eqnarray}
\Psi_{c}^j(z) =  P_j^{-\alpha} \cdot \frac{1}{\sqrt{2\pi}}
\int_{-\infty}^{\infty} F_{c}(q) G_\alpha(k-q) \exp(ikz) dk dq \, ,
\\ \Psi_{v}^j(z) = P_j^{\alpha-1} \cdot \frac{1}{\sqrt{2\pi}}
\int_{-\infty}^{\infty} F_{v}(q) G_{1-\alpha}(k-q) \exp(ikz) dk dq
\,,
\end{eqnarray}
and will satisfy the general BC of Eq. (\ref{Kanedlra}) with
$\alpha_c=\alpha_v=0$.

General BC, for heterostructures described by the MEMA Hamiltonian,
that conserve the normal  component of the envelope flux density
$J_\tau$ have been suggested  by Kisin {\it et al.} \cite{gelmont}.
We have shown that the general requirement  that the heterostructure
MEMA Hamiltonian be self--adjoint leads to the more general
condition $J_\tau^{\alpha \beta}=const$. That is, that $J_\tau$ must
be continuous for arbitrary chosen envelope functions $\Psi_\alpha$,
$\Psi_\beta$. This generalization to $\alpha \ne \beta$ is
important, because it shows that the components of the transfer
matrix  connecting the components of the envelope function $\Psi$
and  envelope normal velocity $V_\tau$ on the two sides of the
interface, are  state independent and are characteristic  properties
of the interface, not< the states.  This is not true for a transfer
matrix which  connects the components of $\Psi$ and its derivatives
$\Psi^{'}$ (see Ref. \onlinecite{gelmont}), because the relationship
between the components of $V_\tau$ and $\Psi$, $\Psi^{'}$, on the
same side of the interface, in the multiband model may depend on the
symmetry (the value of the interface parallel momentum in planar
heterostructures \cite{gelmont} and the total orbital angular
momentum and parity in spherical dots \cite{sercel}) or on the
energy of the state.

An explicit treatment of "interface" effects can be seen in several
${\bf k} \cdot {\bf p}$ models of heterostructures that consider
exact electron wave functions  in the interface region
\cite{foremanburt,einevoll,smith,aversa,burt,foremanmain}. Some of
these models result in equations  in addition to the bulk  ${\bf k}
\cdot {\bf p}$ equations that describe the interface
\cite{burt,foremanmain}. The parameters of , these "interface"
Hamiltonians must be determined from microscopic wave functions or
obtained  from the comparison with experiment. However,  these
advanced models are generally much more complicated than the MEMA,
completed by an appropriate choice of boundary conditions, which has
already successfully described  such  interface effects as  the
$\Gamma - X$ intervalley mixing in $GaAs/AlAs$ heterostructures
\cite{andoII,fu}, the  heavy--light hole  plane wave mixing at zinc
blende interfaces under normal incidence \cite{ivkamros}  and the
effect  on the interband light absorption of the intervalley
conversion of the electron at the surface of indirect--band--gap
semiconductors \cite{brag_prb_98}.

Heterostructures with abrupt heterointefaces, ($L\gg a$),
 are described by the MEMA models using energy
 independent transfer matrices  which characterize the effects of the
 interfaces on the carrier wave functions
  within some energy interval. The energy of the  state considered
   should be within the range of validity of the chosen bulk
MEMA Hamiltonian in each layer. This means, that the energy
should be smaller than the distance to the energy band
extrema not explicitly included in the multiband model.
 In the
appropriate energy interval, the elements of the transfer
matrices can be treated as energy independent trial parameters in
fitting the experimental data.

The number of independent transfer matrices required to ensure the
flux continuity of Eq. (\ref{generalflux}), is determined by the
number of physically nonequivalent  interfaces of the
heterostructure. Therefore, to write general BC, one  must model the
symmetry of the interface, i.e.  the interface geometry and  the
symmetry of the material properties on the interface. On the other
hand, this is true for any explicit determination of the full
microscopic wave function near the interface.

In conclusion, we have derived a general form of
 state--independent BC for multicomponent envelope functions that
are valid under the same conditions as the MEMA itself. Spurious
components of the wave  functions are eliminated  by requiring that
the envelope flux density  be determined to the same order of
approximation as the MEMA itself.  The effect of the general BC on
the electron and hole energy spectra has been demonstrated in the
two band model for a potential well with infinite potential
barriers. This procedure, using the general BC for truncated
envelope wave functions, can be applied  to any MEMA model
containing  any type of spurious solution and  to heterostructures
with finite or infinite potential barriers.

\section*{Acknowledgments}

A. V. Rodina gratefully acknowledges financial support from the
Alexander von Humboldt Foundation. The research of A. Yu.  Alekseev
was supported in part by a grant of the Swedish Research Council
(NFR) F 5102-672/2001 and by INTAS grant INTAS 99-1705. Al. L. Efros
and M. Rosen thank the US Office of Naval Research (ONR) for
financial support.

\begin{thebibliography}{99}
\bibitem{small} The  case, where other small parameters are built into
some elements of the tensor $D_{\mu \, \nu}$ may be treated
in a similar way.
\end{thebibliography}
\end{document}